\documentclass[oldversion,preprint]{aa}  
\usepackage{graphicx}
\usepackage{txfonts}
\begin{document}
   \title{Absolute kinematics of radio source components\\ 
          in the complete S5 polar cap sample}

   \subtitle{III. First wide-field high-precision astrometry at 15.4\,GHz}

   \author{I.~Mart\'{\i}-Vidal\inst{1}
          \and
           J.~M.~Marcaide\inst{1}
          \and
           J.~C.~Guirado\inst{1}
          \and
           M.~A.~P\'erez-Torres\inst{2}
          \and
           E.~Ros\inst{3}
          }

   \offprints{I.~Mart\'{\i}-Vidal}

   \institute{Departament d'Astronomia i Astrof\'{\i}sica, 
              Universitat de Val\`encia, C/ Dr. Moliner 50,
              E-46100 Burjassot (Val\`encia, Spain)\\
              \email{I.Marti-Vidal@uv.es}
         \and
              Instituto de Astrof\'{\i}sica de Andaluc\'{\i}a,
              C/ Camino bajo de Hu\'etor 50, E-18008 Granada (Spain)
         \and
              Max-Planck-Institut f\"ur Radioastronomie, 
              Postfach 2024, D-53010 Bonn (Germany)
             }

   \date{Accepted on 17 September 2007}

% \abstract{}{}{}{}{} 
% 5 {} token are mandatory
 
  \abstract{We report on the first wide-field, high-precision astrometric 
analysis of the 13 extragalactic radio sources of the complete S5 polar cap 
sample at 15.4\,GHz. We describe new algorithms developed to enable the use of 
differenced phase delays in wide-field astrometric observations and discuss 
the impact of using differenced phase delays on the precision of the wide-field 
astrometric analysis. From this global fit, we obtained estimates of the relative 
source positions with precisions ranging from 14 to 200\,$\mu$as at 15.4\,GHz, 
depending on the angular separation of the sources (from $\sim$1.6 to 
$\sim$20.8\,degrees). These precisions are $\sim$10 times higher than the 
achievable precisions using the phase-reference mapping technique.}


\authorrunning{I.~Mart\'{\i}-Vidal et al.}
\titlerunning{Absolute kinematics of the S5 polar cap sample}

   \maketitle
%
%________________________________________________________________

\section{Introduction}

In the past few years, we have carried out a series of very-long-baseline-interferometry (VLBI) observations, using the Very Long Baseline Array (VLBA) at 8.4, 15.4, and 43\,GHz, aimed at studying the absolute kinematics of a complete sample of extragalactic radio sources using astrometric techniques. The target of our programme is the ``complete S5 polar cap sample'', consisting of 13 radio sources from the S5 survey (K\"uhr et al. \cite{Kuhr1981}, Eckart et al.  
\cite{Eckart1986}). All sources in this sample have flux densities higher than 0.2\,Jy at the epoch of our observations and well-defined {\em International Celestial Reference Frame} (ICRF-Ext.2) positions (see Fey et al. \cite{Fey2004}). The relative source separations, less than about 15$^{\circ}$ for neighbouring sources, should allow for typical astrometric precisions in the range of 20 to 100\,$\mu$as, depending on the observing frequency. 
(Lower frequencies translate into lower precisions.) 
Even though the essence of our global differenced phase delay astrometry 
is the same as that of phase-reference astrometry (e.~g., Beasley \& Conway 
\cite{Beasley1995}), there are important differences between them.

On the one hand, the coordinates of one source (the target source) are determined in phase-reference astrometry with respect to the coordinates of another source (the reference source). In our global analysis, we use data from all the 13 sources of the S5 polar cap sample simultaneously, in a unique fit, which increases the precision of the astrometry by a factor $\sim$10, as we will see in Sect.~4. 
On the other hand, phase-reference astrometry requires small angular separations 
between the sources (up to a few degrees). In our global analysis, this requirement is dramatically relaxed. Consequently, we should be able to apply a global differenced phase delay astrometric analysis, similar to the analysis reported here, to a set of sources spread across the whole sky.

We have already described the goals of our astrometric programme in Ros et al. 
(\cite{Ros2001}, hereafter Paper~I) and presented VLBA maps obtained at two 
different epochs at 8.4\,GHz in Paper I and at 15.4\,GHz in P\'erez-Torres et 
al. (\cite{Perez-Torres2004}, hereafter Paper~II). 
In this paper, we present the first conclusive phase delay astrometric results 
of this programme. We describe in Sect. 3 the details of our astrometric analysis 
and our algorithm developed in-house from which we automatically derive the 
2$\pi$$-$ambiguities inherent in the phase delay observable. We also discuss the 
contribution of differenced observables in the astrometric precision for sources 
distributed across a large portion of the sky. 

\section{Observations and maps}

We observed the complete S5 polar cap sample at 15.4\,GHz on 15 June 2000 using all the antennas of the VLBA for 24\,hr. The observations took place in subsets of 3 or 4 radio sources, which were always observed at high elevations. The sources of each subset were observed cyclically for about 2\,hr. 
On-source scans were 60\,sec long, with a small time gap (10-20\,sec) to slew the VLBA antennas. Thus, one complete observing cycle was about 5\,min long.

\begin{figure}
\begin{center}
\includegraphics[width=9cm,angle=0]{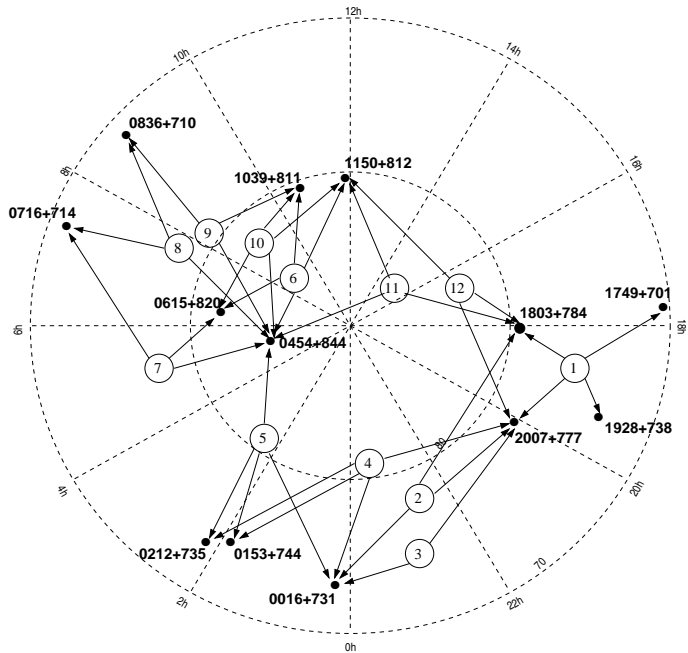}
\caption{Sky distribution of the polar cap sample sources and schematic representation 
of the schedule of our observations. Each circle points to the sources included in the 
same observing block, typically lasting 2$-$4\,hours with a duty cycle of about 
5\,minutes. The numbers associated to the circles represent the chronological order of 
the observation blocks.}
\label{SCHEDULE}
\end{center}    
\end{figure}

This observing mode resulted in a total observation time for each radio source of about 4\,hr (see Fig. \ref{SCHEDULE}). Data were cross-correlated at the Array Operation Center of the National Radio Astronomy Observatory (NRAO). We used the NRAO Astronomical Image Processing System ({\sc aips}) for the calibration of the visibilities. We aligned the visibility phases through the whole frequency band (for all sources and times) by first fringe-fitting the single-band delays of one scan of source 1803+784 and then applying the estimated corrections to all the visibilities. Thus, another fringe-fitting using the multi-band delays provided the new phase corrections for all the observations. 
The visibility amplitude calibration was performed using the system temperatures 
and gain curves from each antenna. For imaging, we transferred the data into the program {\sc difmap} (Shepherd et al. \cite{Shepherd1995}) and made several iterations of phase and gain self-calibration until obtaining high-quality images (with residuals close to the thermal noise). The images of all the sources corresponding to this epoch were analysed and published in Paper~II. The structures of the S5 polar cap sources for our epoch, and other epochs at 15.4\,GHz and 8.4\,GHz, were discussed in Papers~II and I, respectively.

\section{Astrometric analysis}

We performed a differenced astrometry analysis based on the 
phase-connection of the phase delays (e.g. Shapiro et al. 
\cite{Shapiro1979}, Guirado et al.  \cite{Guirado1995}, Ros 
et al. \cite{Ros1999}, and P\'erez-Torres et al. 
\cite{Perez-Torres2000}), but with some substantial differences. 
For the astrometric fits (i.e., estimates of the coordinates of 
the sources via weighted least-square fits of the astrometric 
observables), we used the {\em{University of Valencia Precision 
Astrometry Package}} ({\sc uvpap}), an extensively improved 
version of the {\sc vlbi3} program (Robertson, 
\cite{Robertson1975}; see also Ros et al. \cite{Ros1999} for 
details of the {\sc vlbi3} model). The main improvements of 
{\sc uvpap} permit the use of the JPL ephemeris binary tables and 
compute the relativistic effects of the Solar System bodies using 
the Consensus Model (McCarthy \& Petit \cite{McCarthy2003}). We 
outline the main steps in the astrometric analysis.

\begin{enumerate}

\item{We used {\sc aips} to obtain the group delay, phase delay, and rate from each observation of each of the thirteen sources, after subtracting all the contributions from the structure of the sources, thus referring the phase delays to the phase centre of the images. We discarded the data from the antennas Mauna Kea and St. Croix, because of the high noise in their corresponding fringes.}
 
\item{We predicted the number of phase cycles between consecutive observations of each of the thirteen sources to permit us to ``connect'' the phase delays (e.~g., Shapiro et al. \cite{Shapiro1979}); the computation of the number of phase cycles was performed by comparing the phase delays with the modelled delays obtained from {\sc uvpap}, using a fit of the clock drifts of the VLBA antennas and the tropospheric zenith delays to the group delay and rate data.}

\item{We refined this phase delay connection using an algorithm that imposes the 
nullity of all the closure phases (see Sect.~3.1 for details).}

\item{We computed the differenced phase delays among the sources observed in the same blocks (see Fig. \ref{SCHEDULE}). These differenced delays are largely free of unmodelled effects from the troposphere, ionosphere, and antenna electronics (e.~g., Marcaide et al. \cite{Marcaide1994}).}
 
\item{We estimated the positions of the 13 sources of the S5 polar cap sample via a global weighted least-square analysis of the undifferenced and differenced data.}

\end{enumerate} 

In the last step, we fitted the positions of all the sources with respect to the phase centre of 0454+844 (whose coordinates were fixed parameters in our fit) along with the tropospheric zenith delays (see below) and clock drifts for each antenna. We chose the source 0454+844 as the reference source, not only because of its position (roughly in the geometrical centre of the sky distribution of the sample, minimising the sum of distances to the other sources), but also because it was the most observed source in the schedule, 
being part of most of the observation blocks (see Fig. \ref{SCHEDULE}). Taking this source as reference therefore provides more stability to the astrometric fit.

Regarding the propagation medium, we modelled the tropospheric zenith delay at each station as a piecewise-linear function, with a separation of 6\,hours between nodes, that is, there were five nodes for each antenna. A priori values at the nodes were calculated from local surface temperature, pressure, and humidity measured at each station, based on the Saastamoinen model (Saastamoinen \cite{Saastamoinen1973}). We used the dry Chao mapping function (Chao et al. \cite{Chao1974}) to determine the tropospheric delays at other elevations than the zenith. Changing the mapping function did not alter our results much; for instance, the astrometric corrections of the relative source positions 
(see below) obtained with the use of the Global Mapping Function (Boehm et 
al. \cite{Boehm2006}) differed less than 8\,$\mu$as with respect to those 
obtained using the Chao mapping function. For the ionosphere, we used 
global ionospheric maps at the epoch of our observations derived from GPS 
data and generated by the Center for Orbit Determination in Europe (CODE). 
These maps (in IONEX format) provide the Earth's total electron content 
(TEC), which we transformed into delays over each station using the 
{\sc aips} task {\sc tecor} (e.~g., Walker \& Chatterjee \cite{Walker1999}; 
Ros et al. \cite{Ros2000}). 

\subsection{Phase-connection algorithm}

\begin{figure}
\begin{center}
\includegraphics[width=3.5in,angle=0]{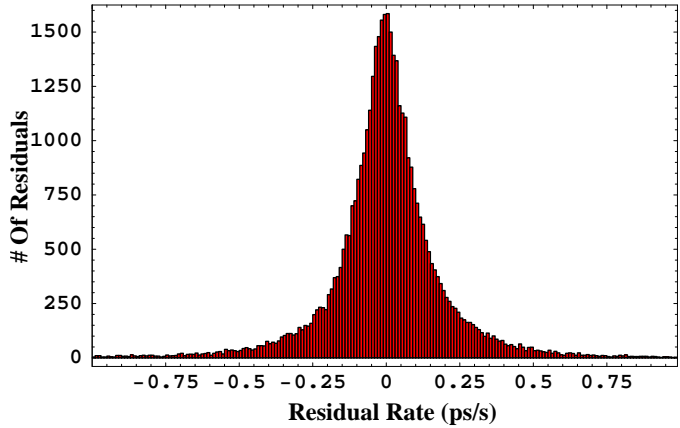}
\caption{Distribution of the residual delay rates for all the baselines, sources, and scans of our observations.}
\label{HISTOGRAM}
\end{center}    
\end{figure}

We used the astrometric model in {\sc uvpap} to predict the number of phase 
cycles ($2\pi$$-$ambiguities) between consecutive observations of the same 
source by means of the residual delay rate. Since the average time separation 
between observations is $\sim$180\,sec and the phase cycle at 15.4\,GHz is 
$\sim$65\,ps, the residual rates have to be lower than (65\,ps/180\,sec)=0.36\,ps/sec 
to ensure a good phase connection. In Fig. \ref{HISTOGRAM} we plot the 
distribution of the residual delay rates of our observations. Residual delay rates 
with absolute values higher than 0.36\,ps/sec in the distribution, most of them 
corresponding to the longest baselines and the weakest sources, are not 
uncommon. Thus, this preliminary phase connection is far from perfect and both 
time- and baseline-dependent phase cycles remain uncorrected in our data. In 
previous works, with less antennas and less sources (e.~g., Guirado et al. 
\cite{Guirado1995}, Ros et al. \cite{Ros1999}), these additional cycles could be 
corrected by inspection of the phase residuals. However, this procedure is 
unmanageable with the amount of data in our observations. Therefore, we have 
devised an algorithm to automate the correction of these unmodelled phase cycles. 
The algorithm works as follows:

\begin{enumerate}

\item For each source and observing time, we compute the closure phases corresponding to all triplets of antennas. From the {\it non-zero} closure phases, we determine the baseline involved most frequently. 

\item We perform an {\em{ambiguity check}} on this baseline. By 
{\em{ambiguity check}}, we mean shifts of the phase delays of that baseline by 
adding or subtracting one phase delay cycle. We perform a {\em positive} (i.~e., 
addition of a cycle) and a {\em negative} (i.~e., subtraction of a cycle) 
{\em shift}. We then compute the {\em scores} corresponding to both {\em shifts}. By {\em score}, we mean the number of closure phases approaching zero minus the number of closure phases distancing from zero after the {\em shift}.

\item We select the {\em shift} with highest {\em score} and modify the data
with such a shift. 

\item We iterate this process (1 to 3). We keep checking on the ambiguities, until all the closure phases (i.~e., closure phase delays) are made zero for the source and time selected. 

\end{enumerate}

We apply this algorithm to all scans in our observations. Actually, to ease the work, for consecutive scans of the same source, we apply the corrections found in the previous scan before performing the {\em{ambiguity check}}.
We have tested this ``automatic connector'' with synthetic data for different scenarios and have obtained excellent results. An example scenario, with a remarkably high noise level, consists of delays equal to a random number of cycles (up to 5) added to randomly selected baselines every random number of scans (with an average of 50 scans between random cycles) in a dataset of 1000 scans and 10 antennas. Under such circumstances, the automatic connector finds 
all the random baseline-dependent cycles without introducing changes in the antenna-based overall constant cycles (see below). We repeated this test several times (with a different number of antennas and even in worse noise conditions) and the connector {\em never} introduced changes in the antenna-based overall cycles. 

\subsection{Antenna-based ambiguities}

The antenna-based ambiguities are offsets of the phase delay, consisting of a given integer number of phase delay cycles that depend on each antenna and source. These ambiguities do not affect the phase closures and, thus, are completely transparent to the automatic connection algorithm described above. 
These antenna-based cycles appear very clearly in the residuals of the differenced delay observables, but can also be detected in the undifferenced residuals. To correct these antenna-based ambiguities, we applied another algorithm based on a {\em smoothness criterion}, which analyses variations between differenced (and also undifferenced) residuals of neighbouring scans that are, in modulus, close to, or larger than, a phase cycle. For each scan, the algorithm

\begin{enumerate}

\item finds these variations. 

\item analyses whether these jumps have an antenna-based structure. 

\item corrects the antenna-based phase cycle in the observations. 

\end{enumerate}

\begin{figure*}
\begin{center}
\includegraphics[width=5.0in,angle=0]{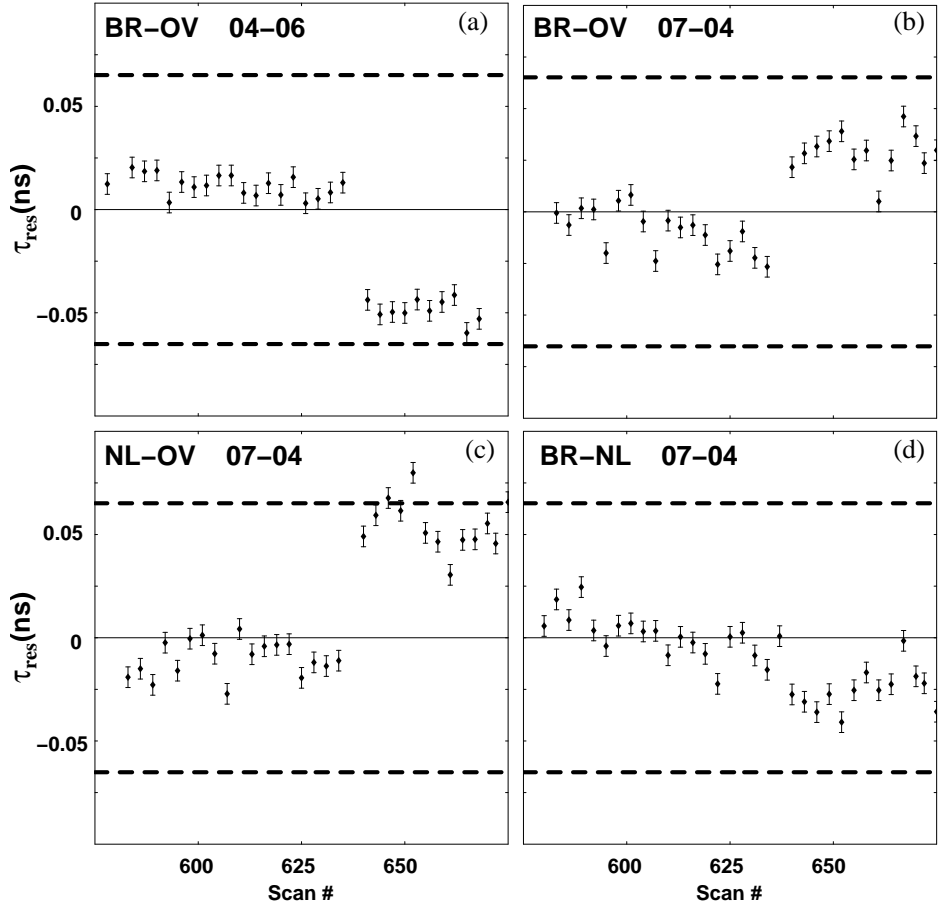}
\caption{Example of the effect of an antenna-dependent unmodelled phase cycle in the differenced observables associated to some baselines and source pairs (see source codes in Table \ref{ASTROMETRY1}). The baseline and source pair corresponding to each figure are indicated in the upper-left corner (BR refers to Brewster, OV to Owens Valley, and NL to North Liberty). The dashed lines represent the delays $\pm$0.06515\,ns, corresponding to $+1$ and $-1$ cycle of phase at the observing frequency. The {\em smoothness criterion} algorithm (see Sect. 3.2) infers, from (a) and (b), that source 04 is the one affected by 
the unmodelled phase cycle and, from (c), that the affected antenna is OV; (d) corroborates the deductions extracted from the other figures.}
\label{SMOOTHER}
\end{center}    
\end{figure*}

This algorithm was applied in a ``bootstrapping'' manner, i.~e., 
beginning with a subset of 3 close-by antennas (Kitt Peak, Pie Town, and Los Alamos) and adding more antennas (one at a time) when all the residuals of the subset of antennas were finally smoothed. We show an example of one iteration of this {\em smoothness criterion} algorithm in Fig. \ref{SMOOTHER}.

\subsection{Overall (source-based) ambiguities}

Since we had phase-connected the data for each source independently, we still had to determine, for each antenna, the overall source-based ambiguity, that is, the integer number of phase cycles by which the phase delay of one source is offset from the others. This offset cannot be totally absorbed by atmospheric, clock, or astrometric corrections, and it can notably affect the astrometric results at our precision level.

We determined the overall ambiguities by again following an iterative process: first, we estimated the overall ambiguities in our weighted least-square fit. The overall ambiguities closest to an integer number of cycles of phase delay were set to be exactly equal to that integer number of cycles of phase delay. Then, we repeated the astrometric fit to obtain new estimates of the 
remaining overall ambiguities. Progressively, all the ambiguities were fixed to integer numbers of cycles of phase delay. For the complete set of overall ambiguities, the maximum deviation with respect to an integer number of cycles turned out to be less than one fourth of a cycle of phase delay (in fact, 40\% of all the overall ambiguities deviated less than one tenth from their closest phase cycle integers).

\subsection{Differenced observables in the global fit}

The use of differenced observables corresponding to a given pair of sources will increase the precision in the determination of the {\em relative} positions of such pair (i.~e., the position of one of the sources with respect to the position of the other). Differenced observables for a total of 24 source pairs (see Table \ref{ASTROMETRY2}) could be computed. This ``network'' of differenced observations introduces redundancies for sources that appear in more than one pair. These constraints in the degrees of freedom for the positions allow for an increase in the precision, not only of the relative source positions, but also of their absolute coordinates, although to a lower degree. The advantage of the redundancy introduced by the network of differenced observations can only be used when more than one pair of sources is available. Thus, the use of all our data in a unique fit provides more robust results than the sub-division of the observations in individual sets of source pairs fitted separately.

Unlike other differenced analyses, in which one source of the pairs was always fixed in the fit, in our global scheme we have differenced observables constructed with pairs of sources whose coordinates are being simultaneously estimated in the astrometric fit (except the coordinates of source 04, which are kept fixed in the fit). This led us to reconsider (see Appendix A) the concept of ``changes in the relative position'' of a pair of sources $(a, b)$, which is now defined as

%\begin{eqnarray}
%\Delta\alpha^{rel}_{a} & = & \Delta\alpha_{b} - \Delta\alpha_{a} -
% \Delta\delta_{a} 
%                             \sin(\alpha_{b}-\alpha_{a}) \tan(\delta_{b})
% \nonumber\\
%\Delta\delta^{rel}_{a} & = & \Delta\delta_{b} - \Delta\delta_{a} 
%                             \cos(\alpha_{b}-\alpha_{a}),
%\label{RelChange}
%\end{eqnarray}

\begin{equation}
\left\{\begin{array}{rl} 
\Delta\alpha^{rel}_{a} & = \Delta\alpha_{b} - \Delta\alpha_{a} - 
\Delta\delta_{a} \sin(\alpha_{b}-\alpha_{a}) \tan(\delta_{b}) \\
\Delta\delta^{rel}_{a} & = \Delta\delta_{b} - \Delta\delta_{a} 
                         \cos(\alpha_{b}-\alpha_{a})
\end{array} \right. ,
\label{RelChange}
\end{equation}

\noindent where $(\Delta\alpha^{rel}_{a}, \Delta\delta^{rel}_{a})$ is the change in the relative position of source $b$ with respect to source $a$, $(\alpha_{a}, \delta_{a})$ and $(\alpha_{b}, \delta_{b})$ are the ICRF-Ext.2 right ascensions and declinations of sources $a$ and $b$, respectively, and $(\Delta\alpha_{a}, \Delta\delta_{a})$ and $(\Delta\alpha_{b}, \Delta\delta_{b})$ are the astrometric corrections for sources $a$ and $b$ resulting from our fit. These expressions have been obtained considering the curvature of the celestial sphere when none of the two sources is kept fixed in our fit. See Appendix A for more details. We notice, in particular, that $\Delta\alpha^{rel}_{b}$ and $\Delta\delta^{rel}_{b}$ need not be the negative of $\Delta\alpha^{rel}_{a}$ and $\Delta\delta^{rel}_{a}$, respectively, since the different orientations of the sources in the sky will produce a combination of their respective changes in right ascension and declination. This is particularly true for the sources close to the celestial pole, as it is the case for the high-declination sources of the S5 polar cap sample.

\subsection{Global phase delay astrometric fit}

We use both differenced and undifferenced observations in the same (global) fit, and the latter are included for the fit to remain sensitive to antenna-dependent 
parameters (i.~e., clock drifts and zenith delays). We scale the standard deviations of the differenced and undifferenced phase delays separately in such a way that, for each type of data and for each baseline and source, the rms of the postfit residuals is unity. The ratio between standard deviations of the differenced and undifferenced observations is $\sim$0.67.

To estimate the astrometric uncertainties, we allowed for variations in the site 
coordinates, the coordinates of the Earth's pole, and UT1$-$UTC in an auxiliary fit, but with their adjustments constrained by their {\it a priori} values and their standard deviations (see Table \ref{IERS-ERROR}) through the use of an {\it a priori} covariance matrix. We then used the final covariance matrix of this auxiliary fit to estimate the final astrometric uncertainties, which now include all the contributions and correlations between the parameters of the geometry of the interferometer and the propagation medium. We also scaled the uncertainties of all the fitted parameters to make the reduced $\chi^2$ equal 
to unity.

Regarding the tropospheric zenith delays, the standard deviations of the fitted nodes of the piece-wise linear functions of our model were $\sim$0.01\,ns. These are smaller than the expected uncertainties due to random variations in the wet component of the tropospheric delay at each site ($\sim$0.1\,ns between nodes; Treuhaft \& Lanyi \cite{Treuhaft1987}). These a priori, extra random variations are not contemplated in the computation of our astrometric uncertainties\footnote{Fixing the uncertainties of the nodes of the tropospheric delay models to 0.1\,ns, would result in an increase in the astrometric uncertainties reported here by a factor of $\sim$2}, because such 
unmodelled variations constitute a large fraction of the final rms (see figure 
\ref{RESIDUALS}), which indirectly affects the astrometric uncertainties after their scaling to obtain a reduced $\chi^2$ equal to unity.

\begin{table}
\caption{Fixed parameters in our astrometric fit and their uncertainties, according to IERS.}
\label{IERS-ERROR}
\centering
\begin{tabular}{c c} 
\hline\hline
{\bf Fixed parameter} & {\bf A-priori uncertainty} \\
\hline
0454+844 position  & 300\,$\mu$as (in $\alpha$ and $\delta$) \\
Earth Pole  & 0.7\,mas (in $\alpha$ and $\delta$) \\
Site Coordinates  & 2\,cm in each coordinate (x,y,z) \\ 
UT1 $-$ UTC  & 0.04\,ms \\
\hline
\end{tabular}
\end{table}

Our global astrometric fit provides the standard deviations of the absolute positions of the sources. Given that we use Eq. \ref{RelChange} to calculate the changes in the relative positions of the sources, the corresponding standard deviations for these changes are given by the following expression:

\begin{equation}
\left\{\begin{array}{rl}
\sigma(\Delta\alpha^{rel}_{a}) & =  \sqrt{\sum_{i,j}^{4}{\frac{\partial\Delta\alpha^{rel}_{a}}
{\partial x_{i}} \frac{\partial\Delta\alpha^{rel}_{a}} {\partial x_{j}} C(x_{i},x_{j})}} \nonumber \\ 
\sigma(\Delta\delta^{rel}_{a}) & =  \sqrt{\sum_{i,j}^{4}{\frac{\partial\Delta\delta^{rel}_{a}}
{\partial x_{i}} \frac{\partial\Delta\delta^{rel}_{a}} {\partial x_{j}} C(x_{i},x_{j})}}
\end{array} \right. ,
\label{Uncertainty}
\end{equation}

\noindent where $x_{i}$ is the set of corrections $(\Delta\alpha_{b}, \Delta\delta_{b}, \Delta\alpha_{a}, \Delta\delta_{a})$, and $C(x_{i}, x_{j})$ is the element $(i, j)$ of the covariance matrix. 

We tested the robustness of our astrometric results by repeating the fit with 
different reference sources. The new absolute source positions were compatible 
with those from the first fit at 1\,$\sigma$ level. The relative source positions 
were compatible at 2\,$\sigma$ level.

\section{Results and discussion}

We present the results of our astrometric analysis in 
Tables \ref{ASTROMETRY1} and \ref{ASTROMETRY2}. 
Table \ref{ASTROMETRY1} shows the astrometric corrections 
and corresponding uncertainties to the absolute source positions 
given by ICRF-Ext.2. Table \ref{ASTROMETRY2} shows: i) the changes in 
the relative coordinates of the 24 source pairs, computed using 
Eq. \ref{RelChange}; ii) the angular separations of the source pairs, 
computed from their ICRF-Ext.2 coordinates; and iii) our estimated 
corrections to the source pair angular separations.

\begin{figure*}
\begin{center}
\includegraphics[width=5.5in,angle=0]{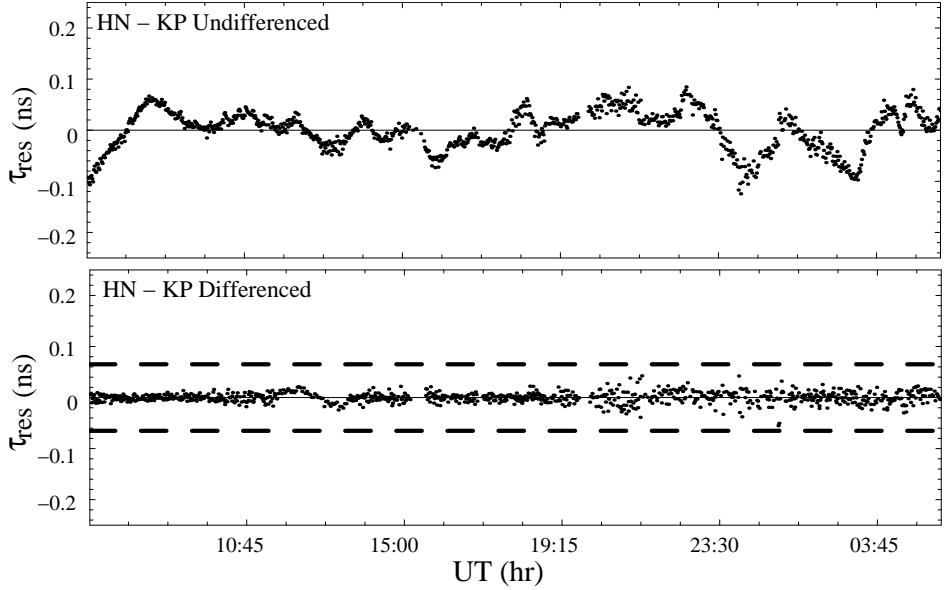}
\caption{Postfit residuals of the undifferenced (above) and differenced (below) phase delays 
of all the pairs of sources for the baseline Hancock $-$ Kitt Peak. The dashed lines below 
represent the delays $\pm$0.06515\,ns, corresponding to $+1$ and $-1$ cycle of phase.}
\label{RESIDUALS}
\end{center}    
\end{figure*}

As an example of the quality of our fit, we show in Fig. \ref{RESIDUALS} the residuals 
of the undifferenced and differenced phase delays corresponding to all the observed sources 
and one of the longest baselines (Hancock $-$ Kitt Peak). Note the cancellation of systematic 
effects in the differenced data, which are still noticeable in the undifferenced data (probably 
unmodelled atmospheric effects). The rms of the undifferenced delays 
for all sources and baselines range between 55\,ps (baseline Fort Davis $-$ North Liberty 
observing the source 02) and 6\,ps (baseline Kitt Peak $-$ Pie Town observing the source 
00). The rms of the differenced delays range from 35\,ps (baseline Hancock $-$ Owens 
Valley observing the pair 11-04) to 2.2\,ps (baseline Brewster $-$ Los Alamos 
observing the pair 18-17). The latter rms is somewhat smaller.

We emphasise that the relative coordinates of each pair in Table \ref{ASTROMETRY2} are not the simple subtractions of the absolute coordinates of the sources forming such a pair in Table \ref{ASTROMETRY1} (except for those pairs with reference to the source 04; see Eq. \ref{RelChange}). Also, the uncertainties in the relative coordinates are much smaller than those of the absolute coordinates. This is a consequence of, first, the natural cancellation of systematic errors in the differenced observables and, second, the correlations between the absolute positions estimated in the fit that account for all posible global shifts of the sources of each pair.

\begin{figure}
\centering
\includegraphics[width=3.5in,angle=0]{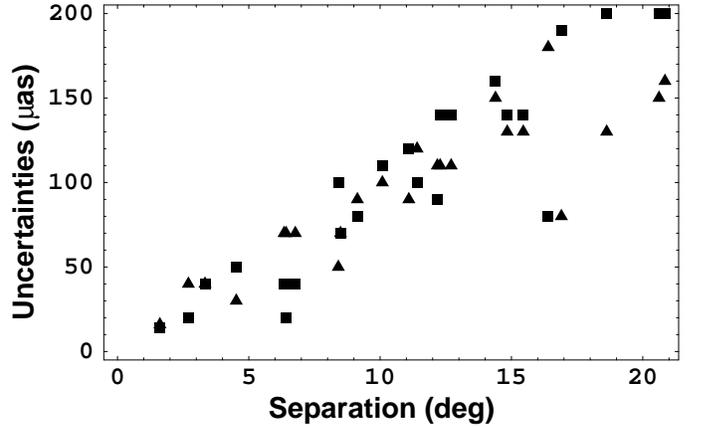}
\caption{Uncertainties in the relative coordinates $\alpha$ (triangles) and $\delta$ 
(squares) of all the pairs of sources as a function of their separations.}
\label{ERROR-DISTANCE}    
\end{figure}

\begin{figure}
\begin{center}
\includegraphics[width=3.5in,angle=0]{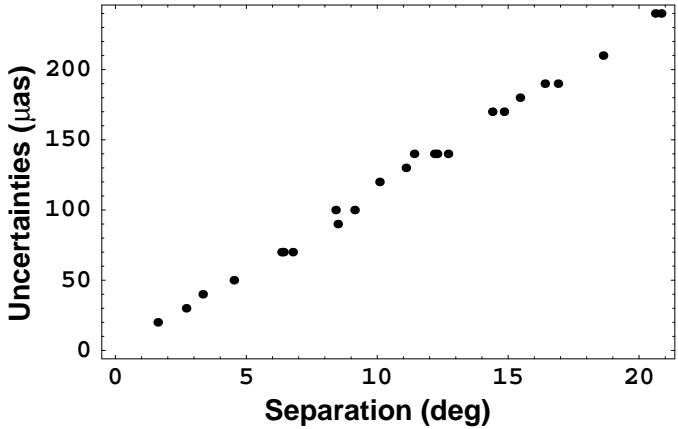}
\caption{Uncertainties in the separations of all the pairs of sources as a function 
of such separations. Since the behavior of the uncertainties is roughly linear, the 
relative errors of the separations of the sources are similar, $\sim$$3\times10^{-9}$.}
\label{ERROR-DISTANCE2}
\end{center}    
\end{figure}

As could be expected, the standard deviations of the absolute positions are roughly 
the same for all sources; however, since we are using differenced observables in 
our fit, the standard deviations of the relative positions of the 24 pairs are 
strongly dependent on the separation of the sources that form the pairs. This effect 
can be clearly seen in Figs. \ref{ERROR-DISTANCE} and \ref{ERROR-DISTANCE2}. Figure 
\ref{ERROR-DISTANCE} shows the uncertainties in the relative $\alpha$ and $\delta$ 
for the 24 pairs as a function of their separations. This behaviour is roughly 
linear, as predicted by the empirical formulae given by Shapiro et al. 
(\cite{Shapiro1979}) and corroborated by simulations of astrometric VLBA 
observations performed by Pradel, Charlot \& Lestrade (\cite{Pradel2006}). This 
linearity is more clearly seen in Fig. \ref{ERROR-DISTANCE2}, which shows the 
uncertainties of the estimates of the separations as a function of such 
separations. However, our astrometric uncertainties are $\sim$10 times smaller 
than those predicted by Shapiro et al. (\cite{Shapiro1979}) and those estimated by 
Pradel, Charlot \& Lestrade (\cite{Pradel2006}) from their simulations, for 
typical declinations of the S5 polar cap sample sources. This improvement in 
precision is probably due to the fact that we have analysed the 13 sources 
simultaneously, instead of using the usual two-source (target/reference) scheme. 
Therefore, our global analysis brings more constraints on the position of the 
sources, which produces an immediate benefit in the precision of the astrometry 
for particularly high-declination sources. Moreover, unlike the phase-reference 
mapping technique, the tropospheric delay model is re-estimated along with the 
source positions in our least-square fit, which also contributes to minimise 
the $\chi^2$.

\begin{table*}
\caption{List of the S5 polar cap sample sources.}
\label{ASTROMETRY1}
\centering
\begin{minipage}{\textwidth}
\begin{tabular}{c c c c c c c c c c} 
\hline\hline
 {\bf Source} & {\bf Alias\footnote[1]{Aliases of source names used throughout the paper.}} & 
{\bf Flux density\footnote[2]{Source flux densities at the epoch of our observations, obtained 
from hybrid mapping using natural weighting.}} & 
{\bf 3$\sigma$\footnote[3]{3 times the root-mean-square of the residual images. By residual 
image, we mean the Fourier transform of the difference between measured visibilities and model 
visibilities.}} & 
\multicolumn{2}{c}{\bf ICRF-Ext.2 position (J2000.0)} & 
\multicolumn{2}{c}{{\bf Astrometric corrections\footnote[4]{Astrometric corrections to the 
ICRF-Ext.2 source positions, resulting from our work. Source 04, fixed at the ICRF-Ext.2 position,
is taken as reference.}}} \\
 & & (mJy) & (mJy/beam) & $\alpha$ & $\delta$ & {\bf $\Delta\alpha$ ($\mu$as)} & {\bf $\Delta\delta$ ($\mu$as)} \\
\hline
0016+731 & 00 &  823.1 & 0.5 & 00$^\textrm{h}$ 19$^\textrm{m}$ 45.786421$^\textrm{s}$ & 73$^{\circ}$ 27$^\prime$ 30.01750$^{\prime \prime}$ & $-$820 $\pm$ 320 & $-$560 $\pm$ 200 \\
0153+744 & 01 &  339.7 & 0.7 & 01$^\textrm{h}$ 57$^\textrm{m}$ 34.964908$^\textrm{s}$ & 74$^{\circ}$ 42$^\prime$ 43.22998$^{\prime \prime}$ & $-$150 $\pm$ 300 & $-$810 $\pm$ 180 \\
0212+735 & 02 & 2519.1 & 1.2 & 02$^\textrm{h}$ 17$^\textrm{m}$ 30.813373$^\textrm{s}$ & 73$^{\circ}$ 49$^\prime$ 32.62179$^{\prime \prime}$ & $-$620 $\pm$ 310 & $-$280 $\pm$ 190 \\
0454+844 & 04 &  225.3 & 0.6 & 05$^\textrm{h}$ 08$^\textrm{m}$ 42.363503$^\textrm{s}$ & 84$^{\circ}$ 32$^\prime$ 04.54402$^{\prime \prime}$ &      0 $\pm$ 300 &      0 $\pm$ 300 \\
0615+820 & 06 &  445.7 & 0.7 & 06$^\textrm{h}$ 26$^\textrm{m}$ 03.006188$^\textrm{s}$ & 82$^{\circ}$ 02$^\prime$ 25.56764$^{\prime \prime}$ &    110 $\pm$ 160 &    200 $\pm$ 100 \\
0716+714 & 07 & 1063.4 & 0.8 & 07$^\textrm{h}$ 21$^\textrm{m}$ 53.448459$^\textrm{s}$ & 71$^{\circ}$ 20$^\prime$ 36.36339$^{\prime \prime}$ &    220 $\pm$ 360 & $-$100 $\pm$ 210 \\
0836+710 & 08 & 1772.9 & 0.8 & 08$^\textrm{h}$ 41$^\textrm{m}$ 24.365262$^\textrm{s}$ & 70$^{\circ}$ 53$^\prime$ 42.17301$^{\prime \prime}$ &    280 $\pm$ 360 & $-$400 $\pm$ 220 \\
1039+811 & 10 &  848.4 & 1.0 & 10$^\textrm{h}$ 44$^\textrm{m}$ 23.062554$^\textrm{s}$ & 80$^{\circ}$ 54$^\prime$ 39.44303$^{\prime \prime}$ &    660 $\pm$ 190 &  $-$30 $\pm$ 110 \\
1150+812 & 11 &  988.3 & 1.0 & 11$^\textrm{h}$ 53$^\textrm{m}$ 12.499130$^\textrm{s}$ & 80$^{\circ}$ 58$^\prime$ 29.15451$^{\prime \prime}$ & $-$450 $\pm$ 180 & $-$640 $\pm$ 120 \\
1749+701 & 17 &  468.7 & 0.8 & 17$^\textrm{h}$ 48$^\textrm{m}$ 32.840231$^\textrm{s}$ & 70$^{\circ}$ 05$^\prime$ 50.76882$^{\prime \prime}$ &    460 $\pm$ 380 &    140 $\pm$ 230 \\
1803+784 & 18 & 2334.5 & 0.7 & 18$^\textrm{h}$ 00$^\textrm{m}$ 45.683914$^\textrm{s}$ & 78$^{\circ}$ 28$^\prime$ 04.01849$^{\prime \prime}$ &     70 $\pm$ 230 &    130 $\pm$ 140 \\
1928+738 & 19 & 3088.6 & 1.1 & 19$^\textrm{h}$ 27$^\textrm{m}$ 48.495167$^\textrm{s}$ & 73$^{\circ}$ 58$^\prime$ 01.57010$^{\prime \prime}$ &     90 $\pm$ 310 &      0 $\pm$ 190 \\
2007+777 & 20 & 1271.0 & 0.5 & 20$^\textrm{h}$ 05$^\textrm{m}$ 30.998511$^\textrm{s}$ & 77$^{\circ}$ 52$^\prime$ 43.24763$^{\prime \prime}$ &    400 $\pm$ 240 & $-$250 $\pm$ 150 \\
\hline
\end{tabular}
\end{minipage}
\end{table*}

\begin{table*}
\caption{Astrometric results for all the observed source pairs.}
\label{ASTROMETRY2}
\centering
\begin{minipage}{\textwidth}
\begin{tabular}{c c c c c c c} 
\hline\hline
{\bf Source pair\footnote{See aliases defined in Table \ref{ASTROMETRY1}.}} & 
\multicolumn{2}{c}{\bf Change in the relative coordinates\footnote{Defined as the change 
in coordinates of the second source with respect to the first one, according to 
Eq. \ref{Rotation}. The uncertainties are estimated using Eq. \ref{Uncertainty}.}} & 
{\bf ICRF-Ext.2 angular separation} & {\bf Correction to angular separation}\\
 & $\Delta\alpha^{rel}$ ($\mu$as) & $\Delta\delta^{rel}$ ($\mu$as) & (deg) & $\mu$as \\
\hline
01-00 &  $-$990 $\pm$ 70 &    180 $\pm$ 40  &  6.770731378 &     850 $\pm$ 70 \\
01-02 &  $-$390 $\pm$ 16  &    520 $\pm$ 14  &  1.614936659 &  $-$620 $\pm$ 20  \\
04-01 &  $-$150 $\pm$ 110 & $-$810 $\pm$ 140 & 12.284604495 &     810 $\pm$ 140 \\
04-02 &  $-$620 $\pm$ 110 & $-$280 $\pm$ 140 & 12.703755162 &     450 $\pm$ 140 \\
04-06 &     110 $\pm$ 40  &    200 $\pm$ 40  &  3.332785200 &  $-$110 $\pm$ 40  \\
04-07 &     220 $\pm$ 150 & $-$100 $\pm$ 160 & 14.393224831 &     150 $\pm$ 170 \\
04-10 &     660 $\pm$ 100 &  $-$30 $\pm$ 110 & 10.085113231 &     390 $\pm$ 120 \\
06-07 &   $-$80 $\pm$ 90 & $-$300 $\pm$ 120 & 11.093608166 &     280 $\pm$ 130 \\
06-10 &     350 $\pm$ 90 & $-$120 $\pm$ 80 &  9.134095175 &     350 $\pm$ 100 \\
08-04 &  $-$400 $\pm$ 180 &    240 $\pm$ 80 & 16.400909237 &     460 $\pm$ 190 \\
08-07 &  $-$180 $\pm$ 70 &    270 $\pm$ 20  &  6.419140221 &     150 $\pm$ 70 \\
08-10 &     730 $\pm$ 110 &    310 $\pm$ 90 & 12.179115583 &     760 $\pm$ 140 \\
11-04 &  $-$350 $\pm$ 120 & $-$120 $\pm$ 100 & 11.407147545 &     350 $\pm$ 140 \\
11-10 &     930 $\pm$ 40  &    580 $\pm$ 20  &  2.699376639 & $-$1020 $\pm$ 30  \\
11-18 &    1280 $\pm$ 130 &    110 $\pm$ 140 & 14.839207812 &     700 $\pm$ 170 \\
11-20 &    1530 $\pm$ 130 & $-$600 $\pm$ 200 & 18.626089045 &    1180 $\pm$ 210 \\
18-00 & $-$1050 $\pm$ 160 & $-$550 $\pm$ 200 & 20.844902015 &  $-$140 $\pm$ 240 \\
18-04 &   $-$60 $\pm$ 80 &     130 $\pm$ 190 & 16.901504141 &   $-$140 $\pm$ 190 \\
18-17 &     340 $\pm$ 50 &     10 $\pm$ 100 &  8.408293010 &   $-$30 $\pm$ 100 \\
18-20 &     260 $\pm$ 70 & $-$360 $\pm$ 40  &  6.342305345 &     370 $\pm$ 70 \\
19-17 &     350 $\pm$ 70 &    140 $\pm$ 70 &   8.491093690 &  $-$360 $\pm$ 90 \\
19-20 &     330 $\pm$ 30  & $-$250 $\pm$ 50  & 4.521891420 &   $-$15 $\pm$ 50 \\
20-00 & $-$1150 $\pm$ 130 & $-$450 $\pm$ 140 & 15.450912042 &  $-$500 $\pm$ 180 \\
20-02 &  $-$910 $\pm$ 150 & $-$300 $\pm$ 200 & 20.618479511 &  $-$300 $\pm$ 240 \\
\hline
\end{tabular}
\end{minipage}
\end{table*}

\subsection{Brief discussion of some selected sources}

All the astrometric corrections shown in Tables \ref{ASTROMETRY1} and
\ref{ASTROMETRY2} are corrections to the ICRF-Ext.2 positions. These ICRF-Ext.2 positions, determined using the group delay observable obtained from dual frequency observations at 8.4 \& 2.3\,GHz, are not directly comparable to our 15.4\,GHz phase delay position estimates: first, the ICRF-Ext.2 positions are not defined with respect to any specific phase centre on the source maps, which our positions are; second, even if the source structure is not astrometrically 
significant, source opacity effects could be present while comparing the source positions determined at 8.4, 2.3, and 15.4\,GHz.

From Table \ref{ASTROMETRY1}, we can calculate mean corrections to the ICRF-Ext.2 coordinates. 
We find mean corrections of 278 and 170\,$\mu$as in right ascension and declination, 
respectively. Our values are similar to those found by Jacobs et al. (\cite{Jacobs2004}). 
These authors are pursuing the extension of the ICRF-Ext.2 to 24 (K-band) and 43\,GHz 
(Q-band). From VLBA global observations they found an agreement between the X/S-frame and 
K-band frame to within 330 and 590 $\mu$as in right ascension and declination, respectively, 
comparable to our mean corrections.

Structure effects must contribute to the corrections of the separations of the source 
pairs given in Table \ref{ASTROMETRY2}. Only seven out of the thirteen S5 sources are 
{\em defining} ICRF-Ext.2 sources and only one of them (source 07) is unresolved 
(following the definition used by Charlot \cite{Charlot1990}) at X-band. Actually, a 
detailed look at the structure of the radio sources can provide some hints to explain 
the relatively large corrections in the last column of Table \ref{ASTROMETRY2}. For 
instance, pairs with source 01 as member, which has an X-band structure index of 4 
(very extended, see Charlot \cite{Charlot1990}) show large corrections to the ICRF-Ext.2 
coordinates; likewise, the corrections corresponding to pairs containing source 11 can 
be explained in terms of a southwest bending of the jet near the core (see maps in 
Paper~I~\&~II). However, a detailed interpretation of the astrometric information obtained 
by comparing with the results from other epochs and frequencies in our astrometric project 
is beyond the scope of this paper and will be published elsewhere. 

\section{Conclusions}

We report here on the first high-precision, wide-field astrometric results at 15.4\,GHz of our multi-frequency monitoring of the S5 polar cap sample. To obtain those results we first developed the package {\sc uvpap}, an extensively improved version of the {\sc vlbi3} program (Robertson \cite{Robertson1975}). The ability of {\sc uvpap} to use differenced phase delays, along with newly developed phase-connection algorithms, enables us the use of differenced phase delays in global astrometric observations. We discuss the impact of the differenced phase delays on our global astrometric analysis and show that their use increases the precision of the relative source positions by a factor of $\sim$10 compared with the precision achievable using the phase-reference technique with a pair of sources (Pradel, Charlot \& Lestrade \cite{Pradel2006}). The astrometric precisions obtained linearly decrease (from 14 to 200\,$\mu$as) as the separations between the sources increase (from $\sim$1.6 to $\sim$20.8\,degrees), with the result that the fractional errors in determining the separations of all the studied source pairs are similar 
($\sim$3$\times10^{-9}$).

We obtain some large corrections for the relative coordinates and separations of 
the sources. From all the 24 pairs studied, 10 have separation corrections above 
500\,$\mu$as and, of those, 4 have corrections above 900\,$\mu$as. These corrections 
could be caused by opacity effects (our observations are at 15.4\,GHz, and the 
ICRF-Ext.2 positions are based on 8.4 and 2.3\,GHz observations) and by source 
structure effects (we are relating our astrometric positions to the phase centres 
of the source maps, that is, the peaks of brightness; the ICRF-Ext.2 positions are 
not well-defined on the source structures). Other wide-field, high-precision 
astrometric analyses of these sources at other frequencies and other epochs are 
currently under way.
This multi-epoch and multi-frequency study will eventually provide spectral 
information and the absolute kinematics for all sources in the sample. Ultimately, 
we expect to provide a definitive test of the stationarity of the innermost 
radio-source cores, associated to the massive black holes, which is a 
basic tenet of the standard jet interaction model (Blandford \& K\"onigl 
\cite{Blandford1979}). In addition, our results will be an excellent complement 
to future $\mu$as-precise astrometry at optical wavelengths.

\begin{acknowledgements}
This work has been partially funded by Grants AYA2004-22045-E, 
AYA2005-08561-C03, and AYA2006-14986-C02 of the Spanish DGICYT. The National 
Radio Astronomy Observatory is a facility of the National Science Foundation 
operated under cooperative agreement with Associated Universities, Inc. We are 
grateful to the anonymous referee for his/her helpful comments and suggestions.
\end{acknowledgements}

\begin{appendix}

\section{Relative position changes for free-moving sources in the sky}

Let $\alpha_{a}$ and $\delta_{a}$ be the right ascension and declination of 
source $a$, and $\alpha_{b}$ and $\delta_{b}$ the right ascension and 
declination of source $b$. Then, the relative position of source $b$ with 
respect to $a$ is

\begin{eqnarray}
\alpha^{rel}_{a} = \alpha_{b} - \alpha_{a} \\
\delta^{rel}_{a} = \delta_{b} - \delta_{a} . \nonumber
\end{eqnarray}

In the simple case that one of the two sources (the reference source $a$) 
is kept fixed in the sky, the change in the relative position between this 
pair of sources can be well defined from equations (A.1): 

\begin{equation}
\left\{\begin{array}{rl} 
\Delta\alpha^{rel}_{a} = \Delta\alpha_{b} \\
\Delta\delta^{rel}_{a} = \Delta\delta_{b} 
\end{array} \right. ,
\label{OldRelChange}
\end{equation}

\noindent where $a$ is the reference (fixed) source and $b$ the target (free) source; $\Delta\alpha_{b}$ and $\Delta\delta_{b}$ are the corrections to the position of source $b$, maintaining the source $a$ fixed in the fit. However, Eqs. (A.1) are not appropriate when the two sources can change their positions in the fit, since the curvature of the celestial coordinate system affects the robustness of Eqs. (A.1) under a global shift of the source pair. To illustrate with a simple example for two sources separated by 12 hours in right 
ascension, a global shift, $\epsilon$, in declination would originate a change of $2\epsilon$ in the relative declination of these sources, according to (A.1). (The declinations of these sources will change with opposite signs.) The relative coordinates between this pair of sources would, then, appear to change dramatically under a global shift of the pair. In other words, we need to define a reference point in the sky to measure the shift of $b$ with respect to $a$. 
In our analysis, for each pair of sources, we select the nominal position of the reference source as the reference point for the study of that particular pair. In practice, such a selection is equivalent to applying a global rotation 
$\Re$ in such a way that the source $a$ is rotated back to its initial (i.~e. nominal) position. According to this rotation, the change in the coordinates of source $a$ is

\begin{equation}
\Re \Rightarrow \left\{ \begin{array}{rl} \alpha_{a} + \Delta\alpha_{a} & \rightarrow \alpha_{a} \\
\delta_{a} + \Delta\delta_{a} & \rightarrow \delta_{a} \end{array} \right. ,
\end{equation}

\noindent where $\Delta\alpha_{a}$ and $\Delta\delta_{a}$ are the corrections to the right ascension and declination of source $a$. This rotation, $\Re$, will have the corresponding effect on the coordinates of source $b$, which are

\begin{equation}
\Re \Rightarrow \left\{ \begin{array}{rl} 
\alpha_{b} + \Delta\alpha_{b} & \rightarrow \alpha_{b} +\Delta\alpha^{m}_{ab} + \Delta\alpha_{b} \\
\delta_{b} + \Delta\delta_{b} & \rightarrow \delta_{b} +\Delta\delta^{m}_{ab} + \Delta\delta_{b} \end{array} \right. ,
\end{equation}

\noindent where $\Delta\alpha^{m}_{ab}$ and $\Delta\delta^{m}_{ab}$ are the changes in right ascension and declination that the rotation $\Re$ causes on the position of source $b$. Thus, the change in the relative coordinates of $b$
with respect to $a$ will be

\begin{equation}
\left\{\begin{array}{rl} 
\Delta\alpha^{rel}_{a} = \Delta\alpha_{b} + \Delta\alpha^{m}_{ab} \\
\Delta\delta^{rel}_{a} = \Delta\delta_{b} + \Delta\delta^{m}_{ab} 
\end{array} \right. .
\label{Rotation}
\end{equation}

When the coordinates of both sources, $a$ and $b$, are corrected in the astrometric fit, we must use these equations instead of Eq. \ref{OldRelChange}.

\begin{figure*}
\centering
\includegraphics[width=4.5in,angle=0]{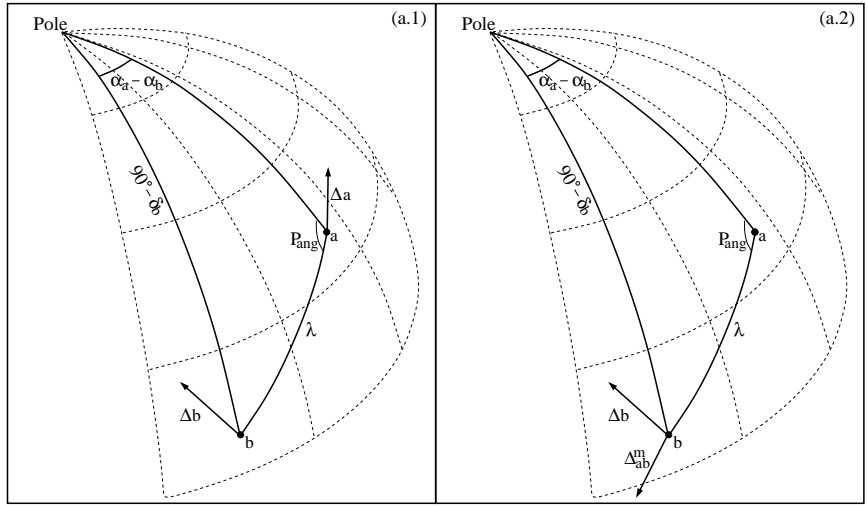}  
\caption{Graphic representation of the rotation $\Re$ that we apply to sources $a$ and $b$. In (a.1), the 
coordinates of both sources, $a$ and $b$, have been corrected an amount 
$\vec{{\Delta}a} = (\Delta\alpha_{a}, \Delta\delta_{a})$ and 
$\vec{{\Delta}b} = (\Delta\alpha_{b}, \Delta\delta_{b})$, respectively. In (a.2), we apply a rotation $\Re$ 
that brings source $a$ back to its ICRF position (maintaining $P_{ang}$ constant) and causes the shift 
$\vec{{\Delta}^{m}_{ab}} = (\Delta\alpha^{m}_{ab}, \Delta\delta^{m}_{ab})$ on source $b$. The total shift 
of $b$ will therefore be the addition of the corrections $\vec{{\Delta}b}$ and $\vec{{\Delta}^{m}_{ab}}$.}
\label{FIG-APPENDIX}  
\end{figure*}

From all possible rotations $\Re$ in the sky that move the source $a$ back to its nominal position, we selected the one that causes the direction of the arc between $a$ and $b$ (i.~e., the position angle of $b$ with respect to $a$) to remain unchanged. According to the sine theorem (see Fig. \ref{FIG-APPENDIX}):

\begin{equation}
\cos{\delta_{b}}\sin{(\alpha_{a}-\alpha_{b})} = \sin{\lambda}\sin{P_{ang}}
\end{equation}

\noindent where $\lambda$ is the arclength between $a$ and $b$, and $P_{ang}$ is the position angle of $b$ with respect to $a$. In our case, this can be expressed as

\begin{eqnarray}
\cos{(\delta_{b}+\Delta\delta_{b})}\sin{(\alpha_{b}-\alpha_{a}+\Delta\alpha_{b}-\Delta\alpha_{a})} = 
\nonumber \\
\cos{(\delta_{b}+\Delta\delta^{m}_{ab}+\Delta\delta_{b})}\sin{(\alpha_{b}-\alpha_{a}
+\Delta\alpha^{m}_{ab}+\Delta\alpha_{b})},
\label{RotTaylor}
\end{eqnarray}

For a first-order approximation of the astrometric corrections ($\sim$0.5\,mas, see Table \ref{ASTROMETRY1}), the condition \ref{RotTaylor}, together with the constancy of the arclength between $a$ and $b$ under the rotation $\Re$, is satisfied if, and only if

\begin{eqnarray}
\Delta\alpha^{m}_{ab} & = & - \Delta\alpha_{a} - \Delta\delta_{a} 
                             \sin(\alpha_{b}-\alpha_{a}) \tan(\delta_{b}) \\
\Delta\delta^{m}_{ab} & = & - \Delta\delta_{a} 
                             \cos(\alpha_{b}-\alpha_{a}). \nonumber
\end{eqnarray}

When we apply these relationships to Eq. \ref{Rotation}, we obtain Eq. \ref{RelChange} of this paper directly:

\begin{eqnarray}
\Delta\alpha^{rel}_{a} & = & \Delta\alpha_{b} - \Delta\alpha_{a} - \Delta\delta_{a} 
                             \sin(\alpha_{b}-\alpha_{a}) \tan(\delta_{b}) \\
\Delta\delta^{rel}_{a} & = & \Delta\delta_{b} - \Delta\delta_{a} 
                             \cos(\alpha_{b}-\alpha_{a}). \nonumber
\end{eqnarray}

\end{appendix}

\end{document}